# Interparticle interactions mediated superspinglass to superferromagnetic transition in Ni-bacterial cellulose aerogel nanocomposites

V. Thiruvengadam and Satish Vitta*
Department of Metallurgical Engineering and Materials Science
Indian Institute of Technology Bombay
Mumbai 400076; India

## ABSTRACT

The interparticle interactions in a magnetic nanocomposite play a dominant role in controlling the phase transitions – superparamagnetic to superspinglass to superferromagnetic. These interactions can be tuned by controlling the size and number density of nanoparticles. Aerogel composites, 0.3 Ni-BC and 0.7 Ni-BC, consisting of Ni nanoparticles distributed in bacterial cellulose have been used as model systems to study the effect of interparticle interactions with increasing fraction of Ni in the composite. Contrary to conventional approach, the size of Ni nanoparticles is not controlled and was allowed to form naturally in the bacterial cellulose template. The structural characterization using x-ray diffraction and electron microscopies indicates the presence of only Ni and cellulose with no other phases. Thermogravimetric analysis shows that the Ni content in the two aerogels is 78 % and 83.8 % respectively. The uncontrolled growth of Ni in cellulose matrix results in the formation of nanoparticles with 3 different size distributions - < 10 nm particles mainly distributed along the length of fibrils, 50 nm particles present in the intermediate spaces between the fibrils and > 100 nm particles present in voids formed by the reticulate structure. The magnetic behavior of these composites as a function of temperature, magnetic field and frequency shows that different magnetic states can be accessed at different temperatures. At room temperature the composites exhibit a weakly ferromagnetic behavior with a coercivity of 40 Oe which increases to 160 Oe at 10 K. The magnetization at both the temperatures however is found to be non-saturating even at fields as high as 20 kOe. The transition from weakly ferromagnetic state at room temperature to a superferromagnetic state at low temperatures is mediated via a superspinglass state at intermediate temperatures. Both the superspinglass state and the superferromagnetic state are found to be sustained by the interparticle interactions aided by the presence of small nanoparticles along the length of cellulose fibres. A temperature dependent microstructural model has been developed combining the structural and magnetic results obtained from the two composite aerogels.

*satish.vitta@iitb.ac.in

# I. INTRODUCTION

The collective behavior of ferro/ferri/antiferro magnetic nanoparticles has been an area of extensive research and continues to be an active area because of its scientific and technological relevance. The sustainability of a domain wall in these materials below a certain size of the particle becomes unviable and hence becomes a single domain particle. The collective behavior of such particle assemblies depends on two main factors:

1. inherent magnetic anisotropy energy of the material, and
2. energy of inter-particle interactions.

The inter-particle interactions can be direct or indirect depending on structural organization of the particles in a medium. At large concentrations if the particles are in physical contact with each other, the collective behavior is due to direct interactions. If the magnetic nanoparticles in such a composite are metallic, it leads to percolation and hence the entire composite exhibits finite electrical conductivity. On the other hand, if the composite is made of insulating magnetic nanoparticles, direct exchange does not lead to finite simple electrical conductivity of the composite. If the particles concentration is reduced to avoid physical contact, then different types of interactions that can influence the collective magnetic behavior are; tunneling, RKKY interaction in the case of a composite with non-magnetic metallic medium separating the magnetic nanoparticles and simple dipole-dipole interactions [1-3]. Hence it can be seen that collective behavior of single domain magnetic nanoparticles is a strong function of their concentration as well as their spatial arrangement. The magnetic behavior can vary from the simple superparamagnetic state to superspinglass and superferromagnetic states as the concentration increases and the interparticle spacing decreases progressively.

These phenomena have been extensively investigated in a variety of materials – bulk as well as thin film forms. Nanoparticles of both magnetic metals and alloys such as Co, Fe, FeNi and FeCo distributed in non-magnetic matrices such as $SiO_2$, $Al_2O_3$, C and Ag as well as ferrite nanoparticles capped with organic surfactant molecules have been used as model systems to study the formation of different magnetic states[4-10]. An ultimate state in the magnetic transitions is the formation of superferromagnetic entities at high concentration of nanoparticles. The formation of superferromagnetic chain like entities has been observed both in ordered as well as disordered arrangement of magnetic particles in a matrix through a systematic increase of concentration/density [11-19] . In all the studies reported so far the major emphasis has been to control the absolute size and size distribution of the nanoparticles such that they remain in a single state of magnetic order and probe the static and dynamic behavior of this state. The natural synthesis processes however result in the formation of nanoparticles with poly-dispersity in the absence of particles growth inhibitors. In such cases different magnetic states can be accessed in a single composite with 'fixed concentration' of magnetic nanoparticles at different temperatures. Magnetic nanoparticles in such a system with different sizes will become unblocked at different temperatures and as the system is cooled progressively it should attain a superferromagnetic state at low temperatures with maximum of interparticle interactions.

In order to investigate the feasibility of above conjecture, Ni-nanoparticles embedded bacterial cellulose composites have been chosen in the present work. Naturally formed bacterial cellulose provides a unique template for the formation of Ni nanoparticles of varying sizes. The long molecular chains together with the fibrous network structure facilitates nucleation and growth of differently sized particles varying from just a few nm to about a hundred nm [20,21]. In the present work aqueous state reduction of $NiCl_2$ in the bacterial cellulose matrix was conducted without growth controlling agents to produce Ni nanoparticles with different sizes. A detailed investigation of the structure as well as static and dynamic magnetic properties of the composites as a function of temperature was performed to identify the different magnetic transitions. At room temperature the composites exhibit large single particles dominated weak ferromagnetic behavior which changes on cooling, first to a superspinglass behavior at intermediate temperatures and then to a superferromagnetic behavior at low temperatures.

## II. EXPERIMENTAL DETAILS

### A. Materials

Food grade sugar, coconut water, orange peel fibers and glacial acetic acid have been used for the synthesis of bacterial cellulose. Pure $NiCl_2.6H_2O$ and $NaBH_4$ procured from Merck chemicals were used for the synthesis of Ni nanoparticles in the bacterial cellulose matrix.

### B. Synthesis of bacterial cellulose

Bacterial cellulose (BC) has been synthesized using an earlier reported technique [22]. The synthesis involves two main steps: i) preparation of inoculums with a large concentration of cellulose synthesizing bacteria, and ii) growth of BC using the inoculums in a rotating disc reactor. The inoculum was prepared first using orange peel fibers and sugar solution. The wild bacteria that grow in this medium produce a thin layer of cellulose on the surface after about a week. This thin layer of cellulose is rich in bacteria and serves as the inoculum to grow large quantities of cellulose in the rotating disc reactor. The reactor consists of several acrylic discs rotating in the nutrient medium –a mixture of coconut water, sugar and glacial acetic acid. The bacteria present in the medium are constantly exposed to air and the nutrient medium and this facilitates cellulose formation on the rotating acrylic discs. After about 15 days sufficient amount of BC gets formed which is harvested for incorporation of Ni nanoparticles, after a thorough cleaning.

### C. Synthesis of Ni-BC composites

Ni nanoparticles are synthesized within the porous structure of BC by aqueous phase reduction of $NiCl_2.6H_2O$ using $NaBH_4$ as the reducing agent. Two separate concentrations of $NiCl_2.6H_2O$ solution, 0.3 M and 0.7 M were used to study the effect of precursor solution concentration on the structure and magnetic properties of the aerogel composites. The composite was synthesized by first immersing BC into the precursor solution for ~ 1 hour after which time the translucent BC becomes green indicating penetration of Ni-solution into the BC matrix. The reduction reaction was then conducted using 2 M concentration $NaBH_4$ under an inert atmosphere of Argon and at ice bath temperature. Inert gas atmosphere is maintained in order to prevent oxidation of the Ni nanoparticles. The green colored BC becomes black due to reduction of $NiCl_2$ into Ni. The reduction reaction was performed for 6 hours to ensure a complete reduction and the as prepared Ni-BC hydrogel was converted to an aerogel by drying the hydrogel in a freezer at – 52°C and $10^{-6}$ bar vacuum for 12 hours. These aerogels, 0.3 Ni-BC and 0.7 Ni-BC were used for all the structural and magnetic studies. The structural characterization was performed by X-ray diffraction with Cu-$K_α$ source, field emission gun scanning electron microscope (FEG-SEM) and high resolution transmission electron microscope (HR-TEM). The magnetic studies were done using a SQIUD – vibrating sample magnetometer (SVSM) at different temperatures and magnetic fields. The magnetic susceptibility as a function of both frequency and temperature was also studied using the SVSM.

## III. RESULTS

### A. Structural characterization

The X-ray diffraction patterns obtained from pristine BC, 0.3Ni-BC and 0.7Ni-BC are shown in Fig.1. The cellulose is known to exist in different polymorphic forms depending on the method of synthesis and the type of bacteria used. The most commonly encountered form in BC however has a triclinic crystal structure belonging to type-I and exhibits peaks in the 2θ range 13°-25°. The presence of the strongest peak at ~ 22° 2θ in all the 3 cases: pristine BC, 0.3Ni-BC and 0.7Ni-BC aerogel clearly shows that BC retains its crystal structure even after being subjected to aqueous state chemical reaction to precipitate Ni-nanoparticles. In addition to the BC peak the 0.3Ni-BC and 0.7Ni-BC aerogels exhibit a broad peak at 45° 2θ corresponding to (111) planes of Ni with no additional peaks. The presence of this peak indicates the crystalline nature of the nanoparticles while the broad nature of this peak shows that the crystallites are extremely small in size. Scanning electron microscopy of pristine BC aerogel reveals the presence of a reticulate structure formed by cellulose microfibers of width 50-75 nm with a wide range of pore sizes, Fig. 2a. The reticulate structure offers pores ranging from nano to meso scales [23,24]. This in fact is evident from the differently sized Ni nanoparticles found in 0.3Ni-BC and 0.7Ni-BC aerogels. Structural investigation by a combination of scanning and transmission electron microscopies, Figures 2b to 2e and Figures 3a to 3d, shows a wide range of variation in the size of Ni-nanoparticles, 5 nm to 250 nm. The particles however can be distinguished into 3 essential size distributions centered around: 10 nm, 50 nm and 100 nm. The small particles~ < 50 nm are in the cavities present along the length of the microfibrils whereas the larger particles ~ > 100 nm are found in the vacant spaces present in the reticulate structure. The number density of nanoparticles centered around 50 nm is found to be the highest with the large nanoparticles present in smallest quantity. This size distribution and number density of the differently sized particles is found to be independent of the concentration of Ni-precursor salt solution indicating that it is a function of the nature of pore size distribution in pristine BC alone, the matrix material. A chemical analysis of the nanoparticles using EDX technique in the scanning electron microscope and selected area diffraction in transmission electron microscope clearly shows that the particles are made of pure Ni with no other phases in the composite, in agreement with the x-ray diffraction results.

The presence of Ni-nanoparticles with different size distribution is a consequence of nano- and meso- porous nature of arrangement of the cellulose fibers. The unique reticulate arrangement has origin in its biogenesis by cellulose producing bacteria. The glucan chain aggregates of thickness 1.5 to 3.5 nm are extruded through the linear pores present on the surface of the bacterial cell wall into ribbons of width 50 – 75 nm. The binary fission and random movement of bacteria in the nutrient medium during cellulose production leads to the formation of a reticulate structure [25] . The space present between the glucan chains in a ribbon and between the ribbons due to their reticulate structure provide nano to meso pores for the formation of Ni-nanoparticles. The extremely small nanoparticles are formed in the pores between the glucan chains while the larger nanoparticles are formed in the void spaces between the cellulose ribbons. The physical environment of space present at different locations limits absolute size of the particles that can form. Since the aqueous state reduction process did not have any nanoparticles growth controlling agents such as surfactants, the nanoparticle size distributions is a replica of the pore structure present in pristine BC.

### B. Thermo gravimetric studies

Thermo gravimetric, TG analysis was performed for pristine BC aerogel and Ni-BC aerogel in the temperature range between 25 and 500°C with a ramp rate of 10 Kmin$^{-1}$ in Nitrogen atmosphere to estimate the amount of Ni present in the two aerogels, Fig. 4. TG curve for pristine BC shows two stage weight loss. First stage weight loss of 6.8% between 50 and 100ºC attributed to evaporation of adsorbed water molecules and the second stage weight loss of ~85% between 300 and 400 °C due to decomposition of BC molecular chains and the remaining 8.2% is the charred residue. TG curves for 0.3Ni-BC and 0.7Ni-BC aerogel show major weight loss of around 10.8% and 7% respectively between 300 and 400 °C. The residual weight other than the charred residue of 8.2% (obtained from TG curve of pristine BC) is a measure of the weight percent of inorganic content present in Ni-BC aerogel which is found to be 78% and 83.8% for 0.3Ni-BC and 0.7Ni-BC aerogels respectively. These results clearly show that increasing the Ni-precursor solution concentration from 0.3 M to 0.7 M results in increasing the Ni content from 78 % to 83. 8 % in the aerogels. In addition, TG curve of both 0.3 and 0.7 Ni-BC aerogels shows a weight gain due to oxidation of Ni nanoparticles at high temperatures.

### C. Magnetic studies

*1. External field dependent magnetization*

The magnetization M of the aerogels as a function of external magnetic field H at different temperatures from 10 K to 300 K, has been studied and is shown in Fig. 5. The magnetization exhibits a hysteretic behavior in both aerogels, 0.3Ni-BC and 0.7Ni-BC, at all the temperatures and does not saturate even at fields as high as 20 KOe. The coercivity H$_C$ increases with decreasing temperature from about 40 Oe at 300 K to 160 Oe at 10 K in both the cases. The magnetization was also found to increase with decreasing temperature for both the aerogels. The magnetization of 0.7Ni-BC in general is found to be slightly higher than that of 0.3Ni-BC, in agreement with the thermogravimetric studies which show a higher Ni content in this aerogel. Although the microstructural studies did not reveal the presence of any phases other than cellulose and Ni, to check the possible presence of either Ni-oxide or Ni-hydroxide which are reported to be antiferromagnetic with T$_N$ ~ 20 K, exchange bias measurements were performed for both the aerogels. The magnetic hysteresis at 10 K under zero field cooled(ZFC) and field cooled (FC), 500 and 1000 Oe, conditions is shown in Figs. 5c and 5d for 0.3Ni-BC and 0.7Ni-BC respectively. The magnetization in both cases, ZFC and FC, is identical and retraces without any shift. These results clearly show that the aerogels consist of BC fibers with pure Ni-nanoparticles distributed in the reticulate structure with no other magnetic phases.

*2. Temperature dependent magnetization*

The variation of magnetization, M with temperature T in both zero field cooled and field cooled conditions has been studied for both the aerogels in the range 5 K to 300 K at different external magnetic fields. The ZFC and FC magnetization curves at lowest external field of 200 Oe for both 0.3Ni-BC and 0.7Ni-BC aerogels are shown in Figs. 6a and 6b respectively. The ZFC magnetization exhibits a typical superparamagnetic like behavior in both cases with a sharp peak at 14.8 K and 17.6 K respectively for 0.3Ni-BC and 0.7Ni-BC aerogels corresponding to the blocking temperature $T_B$. Apart from the sharp peak at low temperature, the magnetization exhibits a shoulder in the range 20 K to 50 K in both cases. In order to confirm that this indeed is the superparamagnetic blocking temperature, the ZFC magnetization was measured at external fields of 500 Oe, 1000 Oe, 2000 Oe and the results are shown in insets of Figs. 6a and 6b. The peak temperature T$_B$ shifts to lower values with increasing magnetic field, a clear signature of superparamagnetic like behavior. The FC magnetization exhibits a monotonic decrease with increasing temperature, a behavior observed commonly in superparamagnetic particles. The FC magnetization however does not merge with ZFC magnetization at $T \sim T_B$, indicating the presence of irreversible magnetization reversal mechanisms in the nanoparticles assembly. In the case of 0.3Ni-BC aerogel the ZFC and FC magnetizations merge at ~ 200 K while for 0.7Ni-BC aerogel it merges above 300 K. This type of behavior is a consequence of the interparticle interactions that play a significant role as well as the broad size distribution of nanoparticles. In order to understand these effects more in detail, ac susceptibility of the two aerogels has also been investigated.

### 3. a.c. susceptibility studies

To understand the nature and dynamics of magnetic interactions between the differently sized Ni-nanoparticles in the aerogels, the ac susceptibility was measured in the frequency range 1 Hz to $10^3$ Hz in the presence of an applied a.c. magnetic field of 2 Oe amplitude in the temperature range 5 K to 70 K, Figs. 7a and 7b. The susceptibility in both the cases, 0.3Ni-BC and 0.7Ni-BC aerogel, exhibits two maxima, $T_{f1}$ and $T_{f2}$ in the temperature intervals 15 K to 20 K and 25 K to 40 K respectively. The low temperature susceptibility maximum $T_{f1}$ has a frequency dependence in the case of 0.3Ni-BC aerogel while it is independent of frequency in the case of 0.7Ni-BC aerogel. The susceptibility maximum $T_{f1}$ shifts to higher values with increasing frequency in 0.3Ni-BC aerogel. The second high temperature broad susceptibility maximum $T_{f2}$ however is found to shift to higher temperature with increasing frequency in both the cases, i.e. 0.3Ni-BC and 0.7Ni-BC. Above 60 K, the susceptibility does not show any frequency dispersion and is independent of frequency. This behavior is typical of supespinglasses where in the interparticle interactions lead to changes in the magnetic relaxation of an ensemble of different sized nanoparticles.

## IV. DISCUSSION

The magnetic behavior of an ensemble of nanoparticles depends on the interparticle interactions apart from the finite size and surface effects. The finite size effects become dominant if the nanoparticles size becomes less than the critical size for the formation of a ferromagnetic single domain particle. The critical size $r_c$ for a ferromagnetic material with weak anisotropy is given by the relation [26];

$$r_c = \sqrt{\frac{9A}{\mu_o M_S^2} \left[ \ln\left(\frac{2r_c}{a}\right) - 1 \right]} \tag{1}$$

where, $A$ is the exchange stiffness constant, $M_S$ the saturation magnetization and $a$ the size of the core taken to be equivalent to the lattice parameter. The critical size for single domain formation for Ni can be determined using equation (1), assuming typical values for the constants [27,28], and it is found to be ~75 nm diameter. This clearly indicates that the smaller particles <50 nm that are present along the microfibers are superparamagnetic where as the larger particles >100 nm found in vacant space between the fibers are ferromagnetic at room temperature. Hence it can be concluded that at any temperature T and external magnetic field H, the specific magnetization M is a result of contributions from both the small superparamagnetic as well as the larger ferromagnetic Ni particles. This behavior is indeed observed in the variation of M as a function of H at room temperature – non-saturating magnetization at large fields with a clear coercivity at low magnetic fields. Hence the overall magnetization is due to a combination of large weakly ferromagnetic Ni nanoparticles of size > 75 nm, small superparamagnetic Ni nanoparticles with size <75 nm and the paramagnetic BC. The resulting magnetization variation of the composites therefore has been fitted using the following equation [29,30],

$$M_T(H) = \frac{2M_S^{ferro}}{\pi} \tan^{-1}\left[\frac{H \pm H_C}{H_C}\tan\left(\frac{\pi S}{2}\right)\right] + M_S^{spm}\left[\coth\left(\frac{\mu^{spm}H}{K_BT}\right) - \left(\frac{\mu^{spm}H}{K_BT}\right)^{-1}\right] + \chi^{BC}H \quad (2)$$

where, $M_S^{ferro}$ is saturation magnetization of ferromagnetic nanoparticles, $S$ the squareness of the loop, $M_S^{spm}$ the saturation magnetization of superparamagnetic nanoparticles, $\mu^{spm}$ the average superparamagnetic moment per particle and $\chi^{BC}$ the susceptibility of paramagnetic BC. The above equation has been used to fit the magnetization hysteresis at different temperatures and the results obtained for the two aerogels, 0.3Ni-BC and 0.7Ni-BC, are given in Table 1. Figures 8(a) and (b) show the fit of equation (2) for the two aerogels. It is seen that the magnetization agrees well with equation (2) confirming the magnetic nature of the Ni nanoparticles present in the aerogel. An interesting point to be noted here is that as the temperature decreases $M_S^{spm}$ increases while $\mu^{spm}$ decreases. This is because, in a system with varying nanoparticles size, i.e, non-monosize distribution, the net magnetization depends both on the number density N of the nanoparticles as well as the effective average magnetic moment of the particles. The average magnetic moment in turn depends on the size 'r' of the nanoparticles at any given temperature. As the temperature decreases, the effective N contributing to M increases while the average size 'r' decreases because smaller and smaller particles moment gets blocked. The contribution to $M_S^{spm}$ therefore from smaller superparamagnetic nanoparticles increases and as a result the average volume of the superparamagnetic particles that contribute decreases. This decrease in average volume of the contributing superparamagnetic particles results in a decrease in the superparamagnetic particles effective average moment. The average magnetic moment of the superparamagnetic particles $\mu^{spm}$ decreases from 3648 $\mu_B$ to 969 $\mu_B$ in the case of 0.3Ni-BC whereas it decreases from 4978 $\mu_B$ to 1310 $\mu_B$ in the case of 0.7Ni-BC when the temperature decreases from 300 K to 100 K. The overall magnetization of the superparamagnetic component in the composite however increases from 0.22 emug$^{-1}$ to 0.31 emug$^{-1}$ in 0.3Ni-BC aerogel while it increases from 0.34 emug$^{-1}$ to 0.45 emug$^{-1}$ in 0.7Ni-BC aerogel on decreasing the temperature to 100 K from room temperature.

The magnetization exhibits coercivity $H_C$ both at room temperature as well as at low temperature, 10 K. Since the small, superparamagnetic nanoparticles magnetic moment is thermally randomized at room temperature, they do not contribute to $H_C$. Hence the coercivity observed at room temperature, 40 Oe and 65 Oe for 0.3 Ni-BC and 0.7 Ni-BC aerogels respectively, is due to the large, multi domain nanoparticles. These values are higher than the typical bulk values and are due to the formation of Ni-nanoparticle chains which increases the effective anisotropy constant for the system. At low temperatures, below $T_B$, however the magnetic moment of superparamagnetic particles is blocked and the $H_C$ observed at 10 K is only due to these small Ni nanoparticles. The temperature dependence of $H_C$ below the blocking temperature $T_B$ for mono-dispersed, non-interacting particles is given by the relation [31], $H_C(T) = H_C(0)\left[1 - (T/T_B)^{1/2}\right]$, where $H_C(0)$ is the maximum coercivity of single domain particles and is given as $0.64 K_1/M_S$. An estimation of the maximum coercivity $H_C$ for the two aerogels at 10 K using the above relation gives 179 Oe and 256 Oe respectively and these are larger than the experimentally determined $H_C$, 157 Oe and 160 Oe respectively. The reduction in $H_C$ compared to the estimated $H_C$ is due to the presence of different sized nanoparticles as well as interparticle interactions in both aerogels. The small difference in $H_C$ between the two aerogels, 0.3Ni-BC and 0.7Ni-BC, clearly shows that the average size of the superparamagnetic particles in the two cases is nearly identical and that only the number density of the particles increases on increasing the precursor solution concentration. An increase in the number density of particles will result in increasing the interparticle interactions. This is further supported by the increased bifurcation between ZFC and FC magnetization observed in 0.7Ni-BC compared to that seen in 0.3Ni-BC, Fig. 6. The progressive freezing of magnetic moments of the interacting nanoparticles increases the irreversibility between ZFC and FC [1,32,33]. The magnetic relaxation in such systems is dominated by the strength of dipolar interparticle interactions $E_d$ and they can exhibit a range of behaviors from superspin glasses to superferromagnets with increasing $E_d$. The dipolar interaction strength, $E_d \propto \mu_{spm}^2/D^3$, where 'D' is the interparticle distance of separation. The high density of nanoparticles present both along the BC fibers and in between the BC fibers leads to extremely small 'D' and hence a large strength of interaction. The presence of interparticle interactions in the aerogels can be inferred from the de-Almeida-Thouless line which separates the superparamagnetic and spin glass phases and is given by the relation [34-36];

$$T_B(H) = T_B(0)\left[1 - BH^{2/3}\right] \qquad (3)$$

Where $T_B(0)$ is the field independent blocking temperature and B a constant which is a function of the nanoparticles magnetic moment vis-à-vis the Curie temperatures. A fit of the magnetic field dependent blocking temperature to equation 3 is shown in Fig. 9a and $T_B(0)$ is found to be 16.5 K and 20 K for 0.3 Ni-BC and 0.7 Ni-BC respectively. These results clearly show that in both the cases the interparticle interactions are strong and that the magnetic behavior is modified by these interactions. This behavior is indeed observed in the ac-susceptibility variation with temperature. Susceptibility temperature maximum $T_f$ in a.c. susceptibility is directly proportional to volume of the superparamagnetic particles through the relation $T_f \alpha (K_{eff} V)/\left[\ln(\tau/\tau_o) K_B\right]$. Therefore, two susceptibility maxima $T_{f1}$ and $T_{f2}$ exhibited in both aerogels, Fig. 7, implies a bimodal size distribution for superparamagnetic Ni particles. The shift in peak temperature of ac-susceptibility $\Delta T_f$ with frequency $f$ defined by the parameter [1,37], $\Phi = \Delta T_f / (T_f \Delta \log f)$ has been found to be 0.007 for $T_{f1}$ in 0.3Ni-BC and 0.04 for $T_{f2}$ in both the aerogels. The extremely small value of $\Phi$ for $T_{f1}$ in the case of 0.3Ni-BC clearly indicates the presence of strong dipolar interactions which can stabilize the superferromagnetic phase. The observation of frequency independent $T_{f1}$ in the case of 0.7Ni-BC aerogel in fact confirms the formation of a superferromagnetic phase at low temperatures due to interparticle interactions. Although the value of $\Phi$ for $T_{f2}$ in both the aerogels is higher compared to that for $T_{f1}$, it is far lower compared to the typical values for a non-interactingsuperparamagnet, > 0.13, indicating that significant interactions are present even at high temperatures. It is to be noted here that coupling of larger, weakly ferromagnetic particles moment occurs with different interaction strengths at different temperatures due to variation in particles size. At $T > T_{f2}$ the interactions are only between larger particles as the magnetic moment of all smaller particles which are present between them is randomized by the large thermal energies. For $T_{f1} < T < T_{f2}$ the magnetic moment of only a fraction of the particles lying between the larger particles gets blocked with their directionality dictated by the larger moments. At $T < T_{f1}$ all the particles magnetic moments are blocked along the direction of pre-existing magnetic moments and this leads to formation of chain like magnetic structures with strong interactions between all the particles. The relaxation dynamics of magnetic moments at the lower temperature $T_{f1}$ in the case of 0.3Ni-BC can only be understood in terms of the critical slowing model for strongly interacting nanoparticles. The relaxation time $\tau$ diverges at the glass transition temperature and is given by the relation [38-42];

$$\tau = \tau^* \left( \frac{T_f^{max}}{T_g} - 1 \right)^{-zv} \quad (4)$$

Where $\tau = (2\pi f)^{-1}$ is the relaxation time corresponding to frequency f, $\tau^*$ the characteristic relaxation time of individual particle magnetic moment, $T_f^{max}$ the frequency dependent susceptibility maximum temperature, $T_g$ the static magnetic freezing temperature and $zv$ the dynamic critical exponent. The frequency dependence of susceptibility maximum $T_{f1}$ in the case of 0.3Ni-BC follows this behavior Fig. 9b with a $T_g$ of 16.5 K and a critical exponent of 7.9. This value of $T_g$ is in agreement with dc magnetization, Fig. 9a, and the dynamical critical exponent is in the range, 4-12 predicted for superspin glass systems [1,2,8,39,42,43].

The absolute strength of interparticle interactions decreases with an increase in temperature and hence the relaxation dynamics of the magnetic moments also changes. The high temperature dispersion at $T_{f2}$, observed in both aerogels, has been found to follow the Vogel-Fulcher behavior corresponding to weakly interacting nanoparticles. In these systems the energy barrier for magnetic moment relaxation is modified by the interaction energy and thus the relaxation time. The relaxation time τ in this case is given by the relation[38,40,41];

$$\tau = \tau_o \exp\left[\frac{E_a}{K_B\left(T_f^{max} - T_o\right)}\right] \qquad (5)$$

where $\tau_o$ is the characteristic relaxation time, $E_a$ the anisotropy dependent activation energy and $T_o$ the strength of interaction between particles. A fit to the susceptibility maximum $T_{f2}$ in both the aerogels, Fig. 9c, gives 154 K and 160 K for $E_a$ and 18.2 K and 22.7 K for $T_o$ for 0.3Ni-BC and 0.7Ni-BC respectively. These results clearly show that the strength of interparticle interaction $T_o$ is not small although it is small compared to the total energy barrier $E_a$, which is largely due to the intrinsic anisotropy energy of the material.

In summary, the different structural studies together with the detailed magnetic studies have been combined to arrive at a temperature dependent magnetic microstructural model of the Ni-BC composite aerogels as shown in Fig. 10. The structural studies indicate 3 distinct size ranges of Ni-nanoparticles in the aerogels – large fraction of extremely fine nanoparticles of size < 10 nm, another fraction of nanoparticles in the range 10 nm – 50 nm and small fraction of particles of size >100 nm. The small Ni nanoparticles are essentially located along the length of the BC fibers while the larger particles are mainly in spaces between the BC fibers formed due to the reticulate structure [Fig. 10(a)]. At temperatures $T > T_{f2}$, all the nanoparticles with size less than the critical size for the formation of single domain particles, 75 nm, are in unblocked state with random alignment of their magnetic moments. The magnetic behavior at these high temperatures is essentially due to the large particles which exhibit a weak ferromagnetic behavior [Fig. 10(b)]. On cooling the nanocomposite to lower temperatures, $T_{f1} < T < T_{f2}$, the magnetic moments of medium sized particles gets progressively blocked with the directionality of the magnetic moments governed largely by the magnetic moments of the larger ferromagnetic nanoparticles [Fig. 10(c)]. As the composite aerogels are cooled further to $T < T_{f1}$, the magnetic moment of even the extremely fine nanoparticles becomes blocked with the moments aligned along directions determined by the larger blocked superparamagnetic particles and the ferromagnetic particles [Fig. 10(d)]. This results in the creation of superferromagnetic entities in the composite formed mainly due to strong interparticle interactions alone. The reducing magnetic distance of separation between the particles as the temperature is decreased leads to coupling and increased interparticles interaction. The formation of chain like superferromagnetic entities in the aerogels is similar to the formation of ferromagnetic chains in ferrofluids wherein the chains are formed due to alignment of nearly monodispersed ferromagnetic nanoparticles in a fluid medium in the presence of an external magnetic field. The surface of BC fibers in the present case promotes preferential nucleation of Ni-particles and thus acts as a template for the formation of chain like superferromagnetic entities. The appearance of a frequency independent peak in the ac-susceptibility of 0.7Ni-BC aerogel clearly supports the presence of superferromagnetic entities in the composite. A measure of the electrical resistivity of the aerogels shows that they are insulators, clearly indicating that formation of superferromagnetic entities is due to indirect interactions such as dipolar interactions and not a result of direct exchange interactions between the nanoparticles.

## V. CONCLUSIONS

The collective behavior of magnetic nanoparticles distributed in an insulating non-magnetic matrix depends on the interparticle interactions. These interactions however are a strong function of the size, size distribution and spatial distribution of the nanoparticles. Hence the collective behavior of magnetic nanoparticles with different strengths of interaction are investigated in magnetic composite aerogels with Ni-nanoparticles distributed in bacterial cellulose matrix synthesized using a simple salt reduction technique. The Ni-particles that form have different sizes varying from ~ 3 nm to 200 nm with distinct distributions. This leads to a unique magnetic behavior depending on temperature. At high temperatures the magnetic behavior is dominated by the large, multi domain Ni-particles which are separated by the smaller superparamagnetic particles. As the temperature decreases, the magnetic moment of the smaller particles becomes frozen with their direction oriented along the dipolar fields of the larger particles. This in turn increases the interparticle interactions and the system behaves as a superspin glass. On further decreasing the temperature, all the nanoparticles magnetic moment becomes completely blocked with the direction of the moments influenced by the prior blocked magnetic moments. This results in the formation of superferromagntic domains which are akin to chains of nanoparticles in a ferrofluid. The BC fibers form a template for this chain like structure with the nanoparticles along their surface connecting the large particles. This type of structure has not been reported earlier and exhibits a superferromagnetic behavior in an aerogel synthesized by a simple technique. Most interestingly, such composite structures with a natural distribution of particles sizes enables a study of magnetic transitions at different temperatures in a single composite as opposed to several composites with controlled sizes required to investigate the different magnetic phase transitions. Since the magnetic nanoparticles are distributed in the cellulose matrix, it can be formed into a magnetic paper with different magnetization dynamics at different temperatures.

**Acknowledgements:** The authors acknowledge Central Facilities of IIT Bombay for providing access to all the experimental facilities.

**Table 1.** The different magnetization parameters obtained by fitting the hysteresis to equation (2). $\mu^{spm}$ is average superparamagnetic moment per particle, $M_S^{ferro}$ is saturation magnetization of ferromagnetic nanoparticles, $M_S^{spm}$ the saturation magnetization of superparamagnetic nanoparticles, $\chi^{BC}$ is susceptibility of paramagnetic BC, $S$ the squareness of the loop.

| Fitting parameters | 0.3 Ni-BC | | 0.7 Ni-BC | |
|---|---|---|---|---|
| | 300 K | 100 K | 300 K | 100 K |
| $\mu^{spm}$ ($\mu_B$) | 3648 | 969 | 4978 | 1310 |
| $M_S^{ferro}$ (emu/g) | 0.38 | 0.51 | 0.53 | 0.7 |
| $M_S^{spm}$ (emu/g) | 0.22 | 0.31 | 0.34 | 0.45 |
| $\chi^{BC}$ (emu/g/Oe) | 2.3×10⁻⁵ | 3.6×10⁻⁵ | 2.7×10⁻⁵ | 3.4×10⁻⁵ |
| $S$ | 0.1 | 0.2 | 0.2 | 0.3 |

**Figure captions**

Fig.1. X-ray-diffraction patterns from pristine BC and the two Ni containing aerogels. Peaks corresponding to triclinic BC can be seen in all the 3 cases with an additional broad peak for Ni in the two aerogel composites.

Fig.2. Scanning electron micrograph of pristine BC (a), shows a reticulate structure formed by flat ribbon like cellulose fibers of width ~ 50-75 nm. The two aerogels, 0.3 Ni-BC and 0.7 Ni-BC, have spherical Ni nano particles of size $\geq$100 nm in voids between the BC fibers (b) and (c), and nano particles of size < 40 nm in the cavities along the BC fibers (d) and (e). EDX of both aerogels show only Ni, C and O peaks (f).

Fig.3. Bright field transmission electron micrographs from 0.3 and 0.7 Ni-BC aerogels show both large, 75 nm to 100 nm (a) and (b) and small, < 10 nm (c) and (d) nanoparticles of Ni. The selected area diffraction pattern in the inset (c) and (d) clearly shows rings only from pure Ni.

Fig.4. Thermogravimetric analysis of pristine BC, 0.3 Ni-BC and 0.7 Ni-BC aerogels indicate weight loss at different temperatures. The initial weight loss below 200°C is due to adsorbed water while for T>200°C the weight loss is due to cellulose charring.

Fig.5. Isothermal field dependent specific magnetization M of 0.3 Ni-BC (a) and 0.7 Ni-BC (b) shows hysteresis at low fields and a non-saturating M at high fields. Insets in (a) and (b) show a magnified view of low field hysteresis. The M-H loops performed both in ZFC and FC conditions at 10K, (c) and (d), to detect exchange-bias. Identical loops without any shift indicate an absence of exchange-bias.

Fig.6. Temperature dependent ZFC magnetization exhibits a maximum at 14.8 K and 17.6 K for 0.3 Ni-BC (a) and 0.7 Ni-BC (b) respectively and a shoulder in the range 20 K to 50 K. Field cooled, FC magnetization decreases with increasing temperature in both cases with large irreversibility. Inset shows field dependent shift of blocking temperature in both the aerogel composites.

Fig.7. The temperature dependent a.c. susceptibility of 0.3Ni-BC (a) and 0.7 Ni-BC (b) exhibits two maxima at $T_{f1}$ and $T_{f2}$. The position of the maximum, $T_{f2}$ shifts to higher temperature with an increase in frequency in both aerogel composites. $T_{f1}$ in the case of 0.7Ni-BC is independent of frequency while it is dependent on frequency in 0.3Ni-BC aerogel composite.

Fig.8. The field dependence of room temperature magnetization has been fitted to equation (1) and is shown by a line through the data points for both the aerogel composites. The magnetization is a sum of contributions from the 3 phases and the inset shows the de-convoluted magnetization.

Fig.9. (a) The blocking temperature $T_B$ varies with external magnetic field in both the Ni-BC aerogel composites and this variation follows the de-Almeida – Thouless dependence. (b) The low temperature susceptibility peak $T_{f1}$ has a dispersion in 0.3 Ni-Bc aerogel composite and the dispersion follows critical slowing dynamics. (c) The high temperature a.c. susceptibility

maximum temperature $T_{f2}$ varies with frequency and this variation follows the Vogel-Fulcher law in both the composites. The line through the data points is a fit to different dependencies.

Fig. 10. The non-magnetic and magnetic microstructure, i.e. arrangement of Ni-nanoparticles in BC matrix has been deduced from structural and magnetic characterizations and are shown schematically. (a) Non-magnetic microstructure showing 3 different size distributions of Ni-nanoparticles in the reticulate BC matrix. (b) At high $T \gg T_{f2}$ only large weakly ferromagnetic Ni-nanoparticles are magnetically active while at $T_{f1} < T < T_{f2}$ (c), intermediate sized Ni-nanoparticles present on the surface of BC fibers also become magnetically active. At $T < T_{f1}$ even the smallest Ni-nanoparticles become magnetically active thus completing the formation of magnetic chains (d).

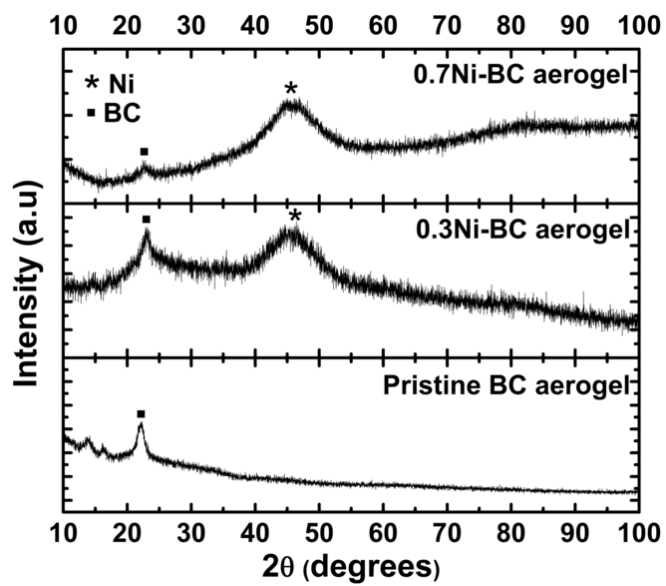

**FIG.1.** X-ray-diffraction patterns from pristine BC and the two Ni containing aerogels. Peaks corresponding to triclinic BC can be seen in all the 3 cases with an additional broad peak for Ni in the two aerogel composites.

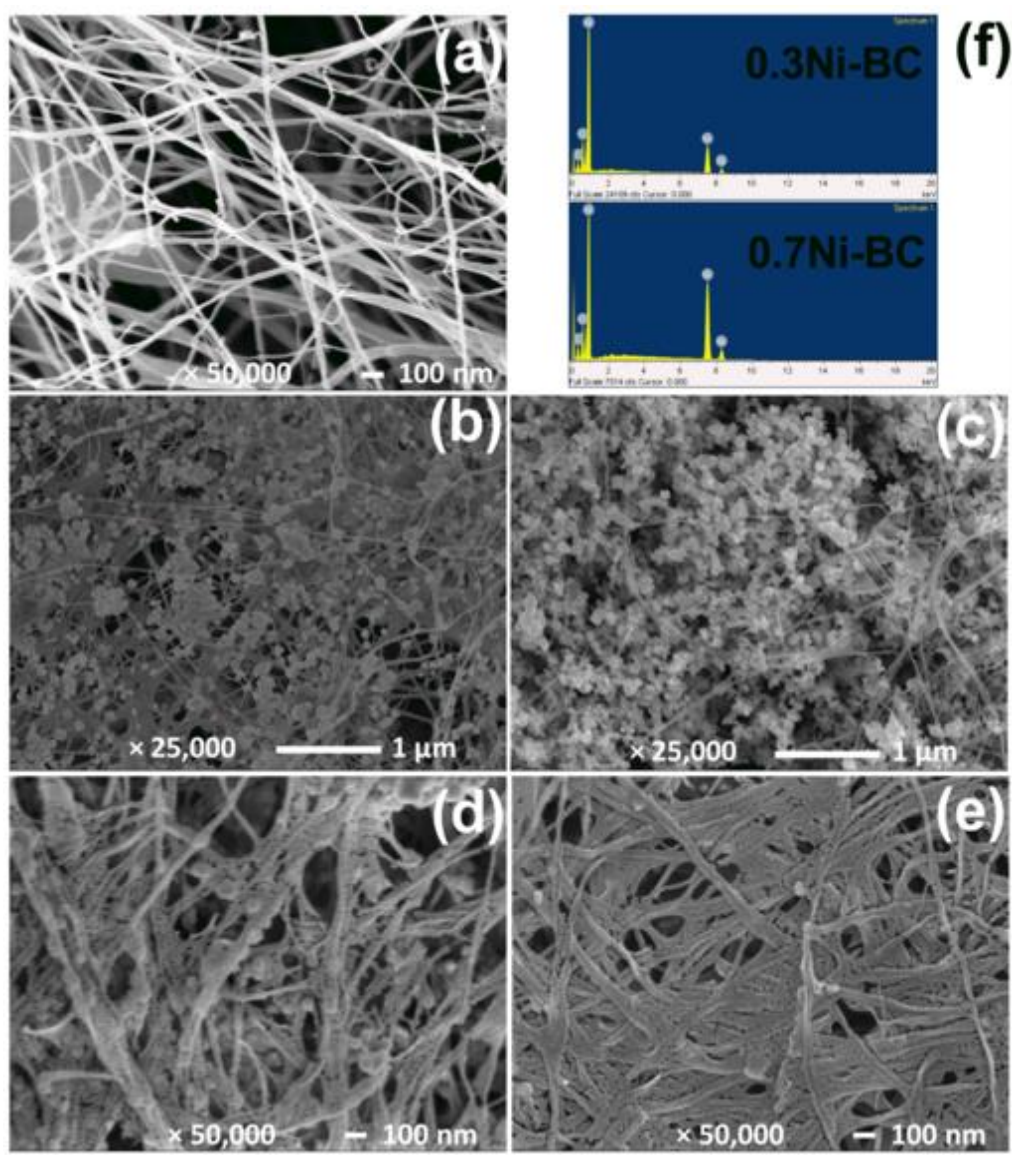

**FIG.2.** Scanning electron micrograph of pristine BC (a), shows a reticulate structure formed by flat ribbon like cellulose fibers of width ~ 50-75 nm. The two aerogels, 0.3 Ni-BC and 0.7 Ni-BC, have spherical Ni nano particles of size ≥100 nm in voids between the BC fibers (b) and (c), and nano particles of size < 40 nm in the cavities along the BC fibers (d) and (e). EDX of both aerogels show only Ni, C and O peaks (f).

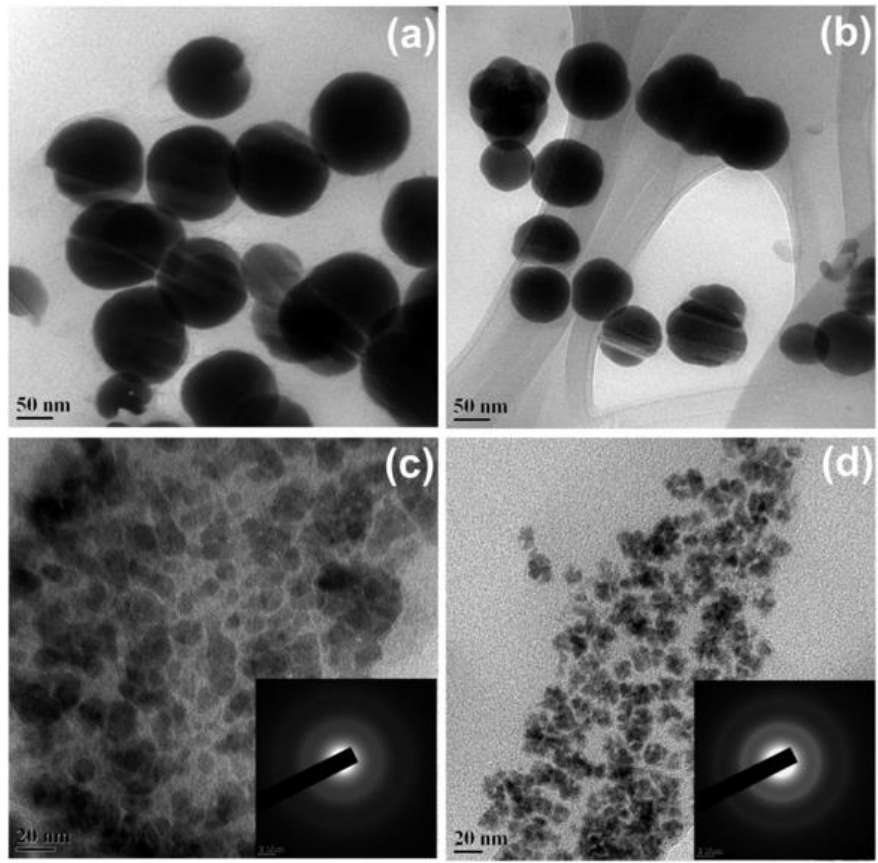

**FIG.3.** Bright field transmission electron micrographs from 0.3 and 0.7 Ni-BC aerogels show both large, 75 nm to 100 nm (a) and (b) and small, < 10 nm (c) and (d) nanoparticles of Ni. The selected area diffraction pattern in the inset (c) and (d) clearly shows rings only from pure Ni.

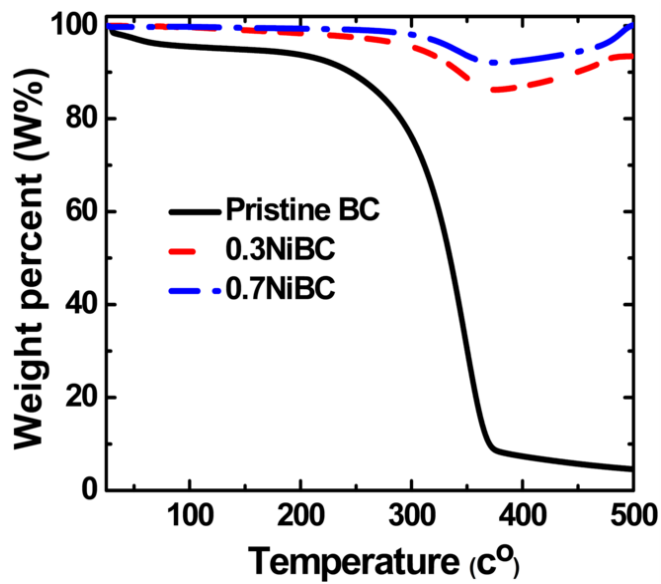

**FIG.4.** Thermogravimetric analysis of pristine BC, 0.3 Ni and 0.7 Ni-BC aerogels indicate weight loss at different temperatures. The initial weight loss below 200°C is due to adsorbed water while for T>200°C the weight loss is due to cellulose charring.

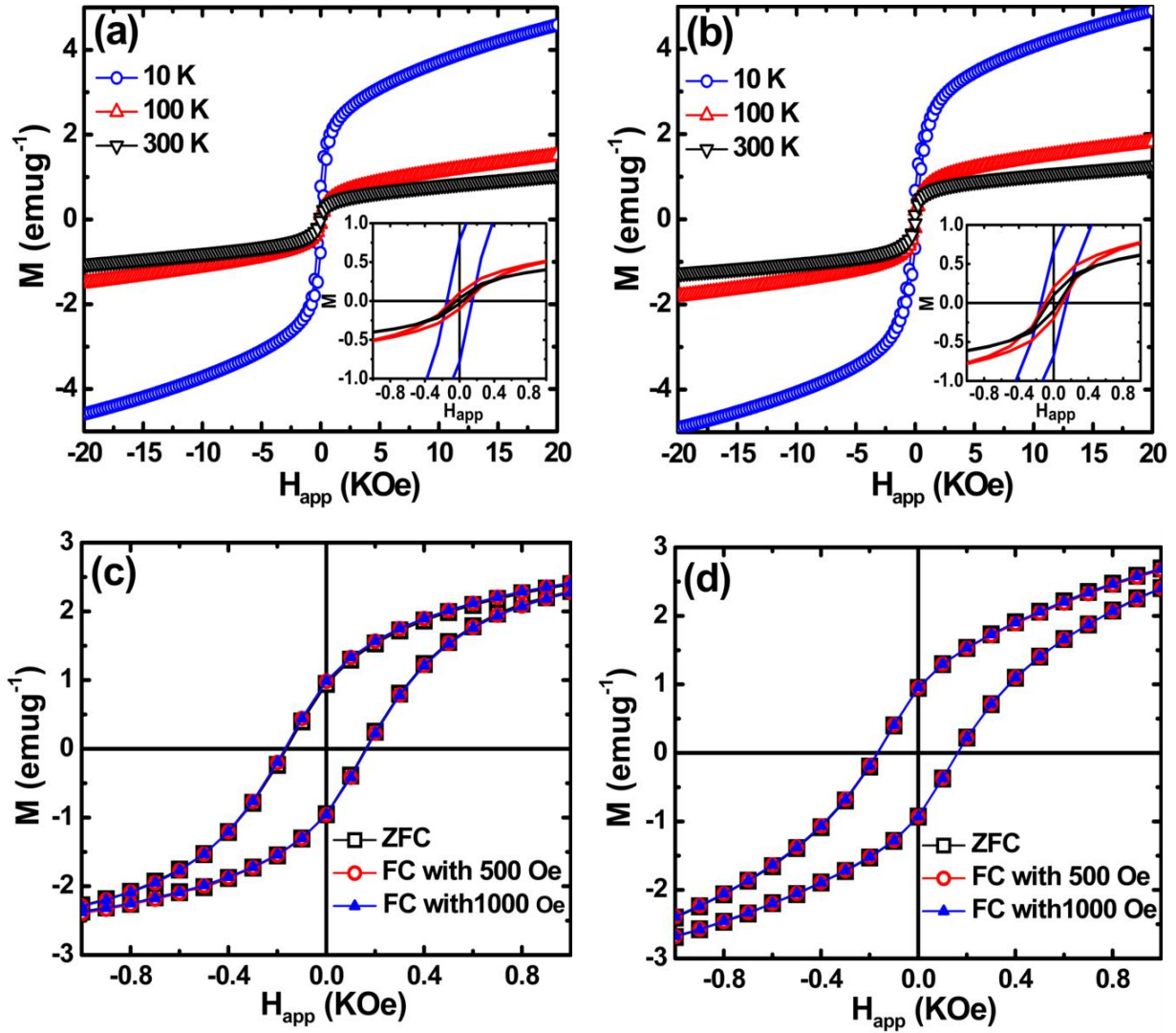

**FIG.5.** Isothermal field dependent specific magnetization M of 0.3 Ni-BC (a) and 0.7 Ni-BC (b) shows hysteresis at low fields and a non-saturating M at high fields. Insets in (a) and (b) show a magnified view of low field hysteresis. The M-H loops performed both in ZFC and FC conditions at 10K, (c) and (d), to detect exchange-bias. Identical loops without any shift indicate an absence of exchange-bias.

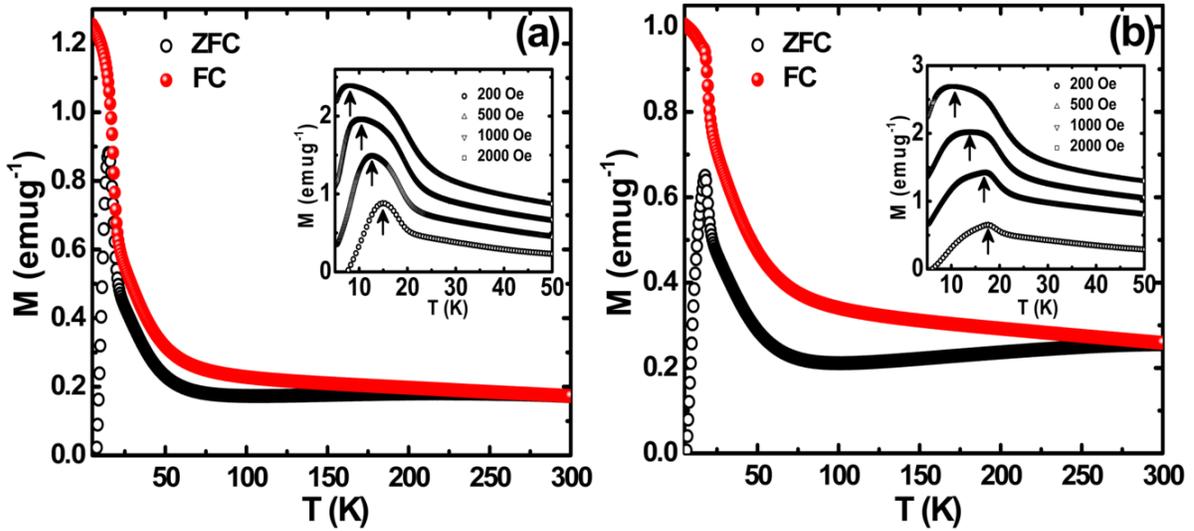

**FIG.6.** Temperature dependent ZFC magnetization exhibits a maximum at 14.8 K and 17.6 K for 0.3 Ni-BC (a) and 0.7 Ni-BC (b) respectively and a shoulder in the range 20 K to 50 K. Field cooled, FC magnetization decreases with increasing temperature in both cases with large irreversibility. Inset shows field dependent shift of blocking temperature in both the aerogel composites.

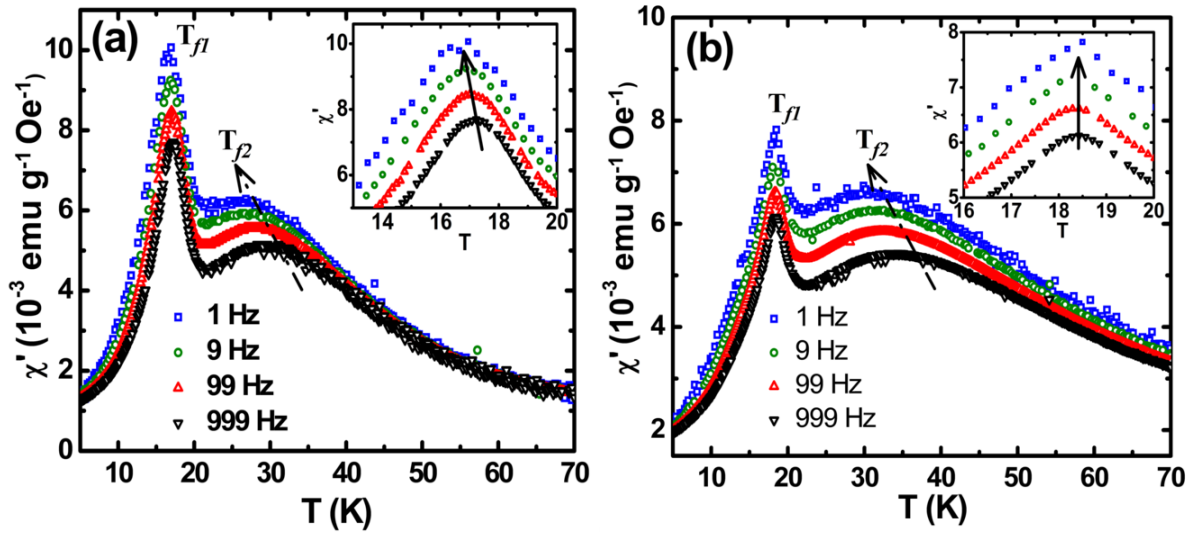

**FIG.7.** The temperature dependent a.c. susceptibility of 0.3Ni-BC (a) and 0.7 Ni-BC (b) exhibits two maxima at $T_{f1}$ and $T_{f2}$. The position of the maximum, $T_{f2}$ shifts to higher temperature with an increase in frequency in both aerogel composites. $T_{f1}$ in the case of 0.7Ni-BC is independent of frequency while it is dependent on frequency in 0.3Ni-BC aerogel composite.

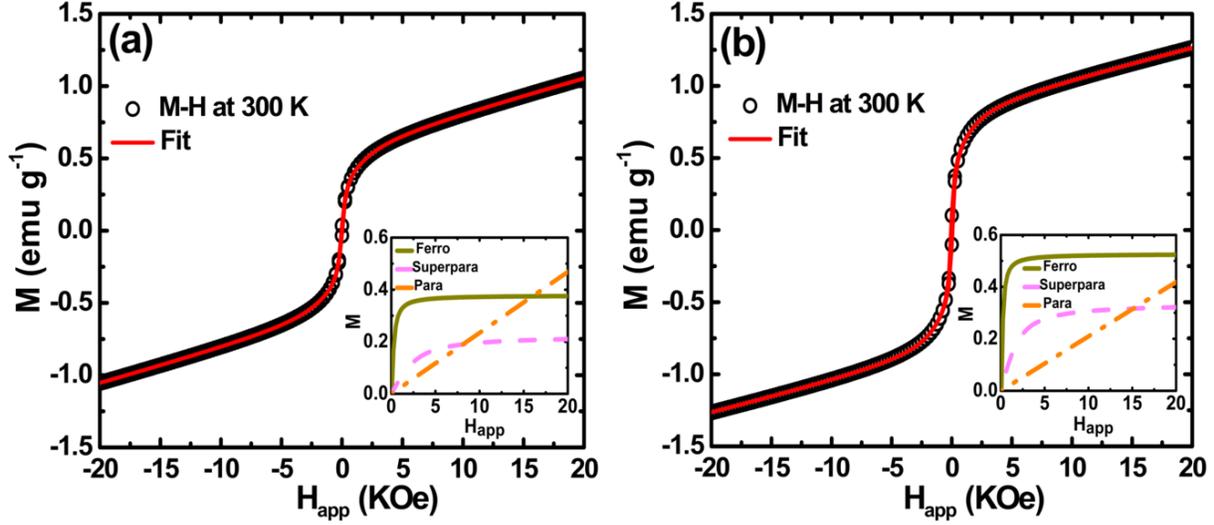

**FIG.8.** The field dependence of room temperature magnetization has been fitted to equation (1) and is shown by a line through the data points for both the aerogel composites. The magnetization is a sum of contributions from the 3 phases and the inset shows the de-convoluted magnetization.

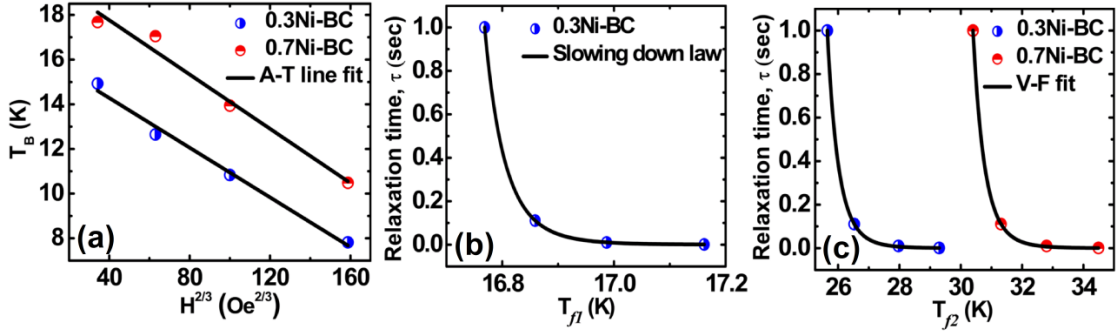

**FIG.9.** (a) The blocking temperature $T_B$ varies with external magnetic field in both the Ni-BC aerogel composites and this variation follows the de-Almeida – Thouless dependence. (b) The low temperature susceptibility peak $T_{f1}$ has a dispersion in 0.3 Ni-Bc aerogel composite and the dispersion follows critical slowing dynamics. (c) The high temperature a.c. susceptibility maximum temperature $T_{f2}$ varies with frequency and this variation follows the Vogel-Fulcher law in both the composites. The line through the data points is a fit to different dependencies.

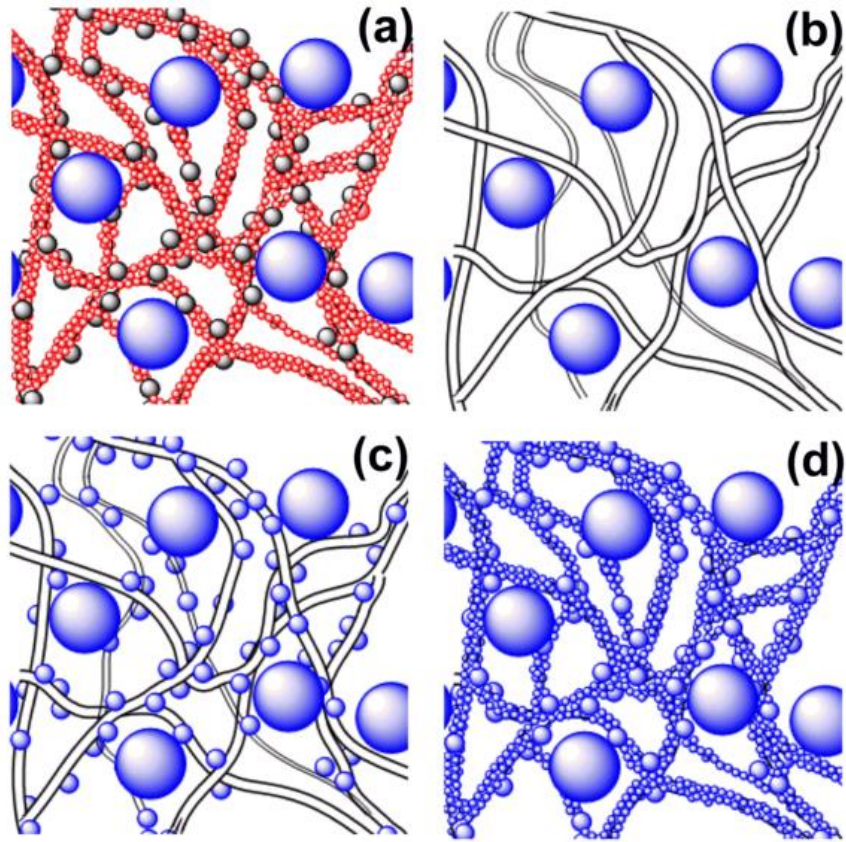

**FIG. 10.** The non-magnetic and magnetic microstructure, i.e. arrangement of Ni-nanoparticles in BC matrix has been deduced from structural and magnetic characterizations and are shown schematically. (a) Non-magnetic microstructure showing 3 different size distributions of Ni-nanoparticles in the reticulate BC matrix. (b) At high $T \gg T_{f2}$ only large weakly ferromagnetic Ni-nanoparticles are magnetically active while at $T_{f1} < T < T_{f2}$ (c), intermediate sized Ni-nanoparticles present on the surface of BC fibers also become magnetically active. At $T < T_{f1}$ even the smallest Ni-nanoparticles become magnetically active thus completing the formation of magnetic chains (d).